\documentclass[aps,twocolumn,preprintnumbers,amsmath,amssymb,superscriptaddress]{revtex4-2}
\UseRawInputEncoding
\usepackage[normalem]{ulem}
\usepackage{placeins}
\usepackage{latexsym,amssymb,amsthm,amsmath,epsfig,braket} 
\usepackage[left=2.00cm, right=2.00cm, top=2.00cm, bottom=2.00cm]{geometry}
\usepackage{xcolor}
\usepackage{svg}
\usepackage[colorlinks, citecolor={blue!80!black}, urlcolor={blue!80!black}, linkcolor={red!50!black}]{hyperref}
\usepackage{booktabs}
\usepackage{subfigure}

\usepackage{hyperref}
\hypersetup{
    citecolor=red,
    colorlinks=true,
    linkcolor=blue,
    filecolor=blue,      
    urlcolor=blue,
}

\begin{document}
\title{Intrinsic Topological Control of the Orbital Hall Effect in Buckled Dirac Materials}

\author{Madiha Zia}
\affiliation{Department of Physics, Quaid-I-Azam University Islamabad, 45320, Pakistan}
\author{Muzamil Shah}
\email{muzamil@qau.edu.pk}
\affiliation{Department of Physics, Quaid-I-Azam University Islamabad, 45320, Pakistan}
\author{Kashif Sabeeh}
\affiliation{Department of Physics, Quaid-I-Azam University Islamabad, 45320, Pakistan}
\author{Gao Xianlong}
\email{gaoxl@zjnu.edu.cn}
\affiliation{Department of Physics, Zhejiang Normal University, Jinhua 321004, P. R. China}
\author{Reza Asgari}
\email{reza4asgari8@gmail.com}
\affiliation{Department of Physics, Zhejiang Normal University, Jinhua 321004, P. R. China}
\affiliation{School of Quantum Physics and Matter, Institute for Research in Fundamental Sciences (IPM), Tehran 19395-5531, Iran}
%\author{Imtiaz Khan}
% \email{ikhan@phys.qau.edu.pk}
%\affiliation{Department of Physics, Zhejiang Normal University, Jinhua, Zhejiang 321004, China}
%\affiliation{Zhejiang Institute of Photoelectronics, Jinhua, Zhejiang 321004, China}
%\author{Kashif Sabeeh}
% \email{ksabeeh@qau.edu.pk}
%\affiliation{Department of Physics, Quaid-I-Azam University Islamabad, 45320, Pakistan}
\date{\today}

\begin{abstract} 

We study the orbital Hall response in buckled two-dimensional Dirac materials using a unified framework that includes an antiferromagnetic exchange field, a perpendicular electric field, and intrinsic spin-orbit coupling. We show that the orbital Hall conductivity is considerably boosted around band-inversion points and shows different signatures across multiple electronic phases using a low-energy massive Dirac model in conjunction with Berry-curvature-based linear response theory. We find a series of quantum spin Hall, valley Hall, and anomalous Hall regimes by methodically adjusting external fields, and demonstrate how the evolution of the orbital response is controlled by the redistribution of Berry curvature between spin and valley sectors. We examine the impacts of finite temperature in more detail and find that although the response's size is suppressed by thermal broadening, the distinctive phase-dependent features remain robust. Our findings demonstrate that orbital Hall conductivity offers a sensitive band topology probe in Dirac systems and emphasize buckled two-dimensional materials as a flexible platform for engineering tunable orbital currents for orbitronic applications.

\end{abstract}

\maketitle

\section{Introduction}

The study of topological phases of matter has transformed our comprehension of electronic transport in condensed matter systems by demonstrating that robust and unconventional responses can arise from the characteristics of Bloch wavefunctions \cite{hasan2010colloquium, qi2011topological, bansil2016colloquium, xiao2021first, konig2007quantum, knez2011evidence, wu2018observation}. The Berry curvature \cite{RevModPhys.82.1959} is a key idea in this framework. It acts like an effective magnetic field in momentum space and controls a number of transverse transport phenomena, such as the anomalous, spin, and valley Hall effects \cite{hasan2010colloquium, qi2011topological, bansil2016colloquium, xiao2021first}. Researchers have looked into these ideas in Dirac materials and two-dimensional topological insulators. In these materials, band inversion and symmetry breaking lead to quantized or almost quantized transport signatures that are linked to nontrivial topological invariants.

Even without an external magnetic field, electrons develop an unusual transverse velocity when subjected to an electric field, attributable to the pivotal influence of Berry curvature in these phenomena \cite{sundaram1999wave, RevModPhys.82.1950}. This physics is realized in two-dimensional systems exhibiting topological phases, such as Chern insulators, which demonstrate chiral edge states and the quantum anomalous Hall effect \cite{haldane1988model, chang2013experimental, deng2020quantum}. In addition to helping scientists find new quantum phases and possible uses in electronics and spintronics, the related Chern invariants give a single picture of these topological insulating phases and show how topology affects both charge and spin transport \cite{yao2026orbital}.

In addition to charge and spin, the orbital degree of freedom has emerged as a significant factor in electronic transport in recent years \cite{RevModPhy.87.1213, go2021orbitronics}. Consequently, orbital transport has become an essential concept in next-generation electronic systems, offering an alternative to conventional charge- and spin-based information processing methods. The orbital Berry curvature of the electronic system governs the orbital Hall effect (OHE), a transverse flow of orbital angular momentum induced by an applied electric field \cite{RevModPhy.87.1213}.

In contrast to spin Hall effect which demands strong spin-orbit coupling, orbital Hall effect (OHE) can be large even for materials with light-elements \cite{RevModPhys.87.1213, vignale2010ten, PhysRevLett.83.1834, tanaka2008intrinsic, PhysRevLett.102.016601, PhysRevLett.121.086602}. Orbitronics can also be viewed as a natural generalization of spintronics \cite{RevModPhys.76.323}, because orbital currents can induce spin currents via spin–orbit coupling \cite{PhysRevLett.102.016601, PhysRevLett.121.086602}. Therefore, controlling the OHE offers a promising way to manipulate spin and information \cite{phong2019optically, cysne2021disentangling, mu2021pure}. Naturally, topological phase transitions serve as a knob to tune in this framework, as they are accompanied by band inversion and orbital changes between the valence and conduction bands \cite{chen2024topology, li2024floquet}. 

The underlying cause of the orbital Hall effect in low-dimensional systems remains unclear despite substantial advancements. Specifically, the degree to which topology influences the OHE as opposed to specific aspects of the band structure, like orbital composition and band gap size, is still unknown. Furthermore, it is yet unclear how altering symmetry and externally adjustable parameters affect the OHE. These elements cause topological phase transitions and have a significant impact on the distribution of Berry curvature in momentum space, which can significantly alter orbital transport properties \cite{jungwirth2012spin, RevModPhys.82.1950}.

The significance of orbital angular momentum transmission in solids has been highlighted by experiments showing large orbital Hall currents in light metals like Ti \cite{choi2023observation}. According to later theoretical research, orbital Hall insulating phases can be achieved by adding symmetry breaking or altering band topology to control the orbital Hall response \cite{ji2024reversal}.   Furthermore, it has been shown that the OHE may be manipulated and topological phase transitions can be induced in two-dimensional ferromagnets using circularly polarized light and Floquet engineering \cite{li2024floquet}. Specifically, Li \textit{et al.} demonstrated that the orbital Hall effect (OHE) can be tuned through topological phase transitions in two-dimensional ferromagnets, where the orbital angular momentum distribution and the resulting orbital Hall response are largely controlled by band inversion \cite{li2026topological}. These results suggest that orbital movement in low-dimensional systems can be effectively controlled by topology. However, the orbital Hall conductivity in minimum two-dimensional Dirac systems is still not well understood in terms of how externally controllable symmetry-breaking fields control it.  More importantly, a unified description of how topological phase transitions influence orbital Hall conductivity in such systems remains incomplete.

A perfect platform to answer these problems is provided by buckled honeycomb materials, including silicene, germanene, and stanene \cite{ezawa2012spin, PhysRevLett.107.076802}. Unlike planar graphene, their intrinsic buckling enables a sublattice potential created by a perpendicular electric field to directly regulate the Dirac mass and related topological phases. Furthermore, magnetic exchange fields can break time-reversal symmetry and further enhance the phase diagram, for instance by being close to magnetic substrates \cite{PhysRevLett.107.076802}. Because of these characteristics, buckled Dirac systems provide an especially well-suited and analytically tractable framework for studying the unified interaction of orbital transport, symmetry breaking, and band topology \cite{RevModPhys.82.1959, PhysRevLett.107.076802}.

\begin{figure}[t]
    \centering
    \includegraphics[width=1.0\linewidth]{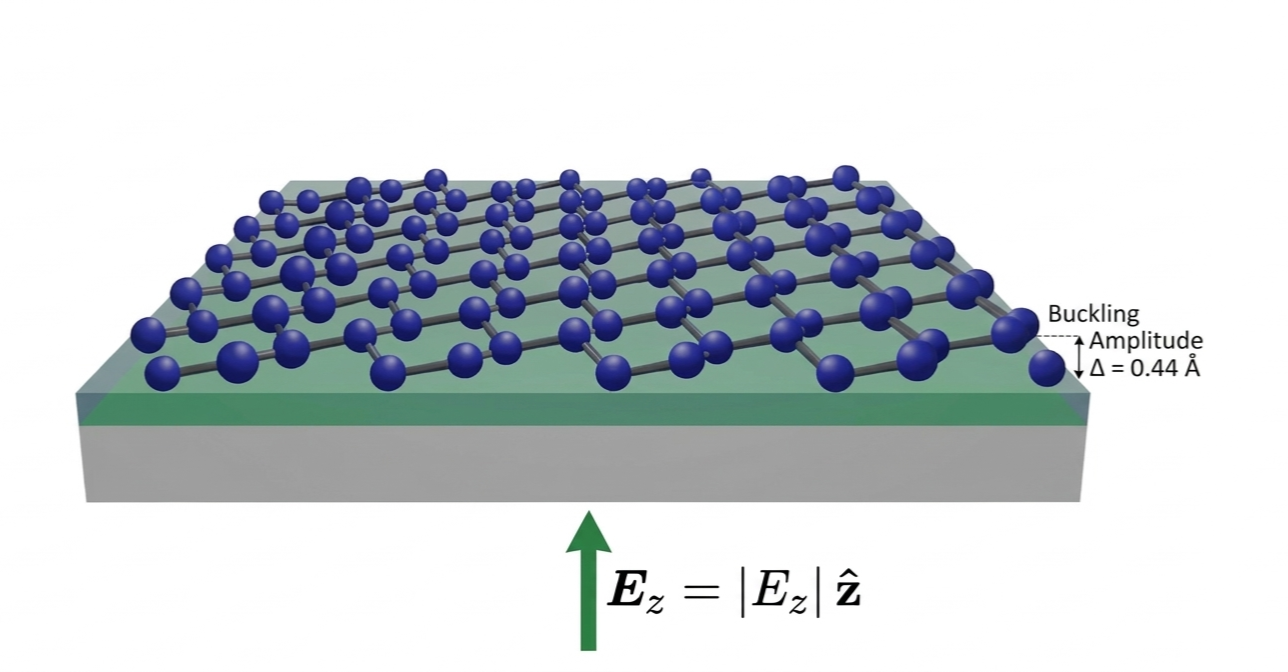}
    \caption{A simple schematic showing a single layer of silicene exposed to an out-of-plane electric field.}
    \label{fig:structure}
\end{figure}

Although orbital Hall effects have been theoretically predicted in multiorbital systems \cite{RevModPhy.87.1213} and transition metals \cite{bhowal2020intrinsic}, a comprehensive understanding of how electric and antiferromagnetic fields jointly shape orbital Hall phases in buckled two-dimensional materials remains incomplete too. In particular, the way symmetry breaking redistributes Berry curvature across the Brillouin zone, as well as the role of finite temperature in modifying orbital Hall conductivity, has not yet been fully clarified.

In this paper, we study the intrinsic orbital Hall effect in a buckled two-dimensional Dirac system under the influence of an antiferromagnetic exchange field and a perpendicular electric field. We examine how the orbital Hall conductivity changes over several electronic phases using a low-energy massive Dirac Hamiltonian and Berry-curvature-based linear response theory. We find distinct signatures related to gap closing and band inversion. We demonstrate that the orbital response is significantly amplified close to topological phase boundaries, where Berry curvature is highly concentrated, and the effective Dirac mass decreases. We clarify the orbital Hall effect's microscopic origin and its dependency on symmetry breaking by resolving contributions from various spin and valley sectors. We investigate the impact of finite temperature further and show that the features of the orbital response persist despite thermal smearing. Therefore, we show that orbital response is enhanced near topological phase transitions, establishing band inversion as a means to manipulate orbital transport in few-layer Dirac systems. 

Our findings show that the orbital Hall conductivity offers a sensitive probe of the underlying band topology and its evolution under external control, even in minimum two-band Dirac models. In addition to expanding our knowledge of orbital transport in low-dimensional quantum materials, these results suggest useful methods for creating adjustable orbital currents in newly developed orbitronic devices.

\section{Theoretical Model}

Buckled two-dimensional honeycomb materials like silicene, germanene, and stanene have garnered special attention among the different material platforms suggested for orbitronic applications. Although the atomic structure of these materials is buckled, they have the same honeycomb lattice shape as graphene~\cite{ezawa2012spin,PhysRevLett.107.076802}. Structural buckling plays a crucial role in shaping the electronic structure and orbital characteristics of the system. To describe the low-energy physics of extended materials without boundary, it is convenient to adopt a valley- and spin-resolved massive Dirac Hamiltonian. Although pristine silicene and germanene do not naturally possess stable long-range magnetic ordering, magnetic exchange fields can still arise in two-dimensional systems through proximity effects in van der Waals heterostructures \cite{zhong2020layer}. Therefore, the antiferromagnetic exchange term considered in the present Hamiltonian is an effective proximity-induced field. Near the Dirac points, the electronic states can be described by the Hamiltonian:

\begin{equation}
\label{silicenehamil}
    {\cal {H}}
=
\hbar v_F\left(\eta\sigma_x k_x+\sigma_y k_y\right)
+\eta s\,\lambda_{SO}\,\sigma_z
-E_z\ell\,\sigma_z
+s\,\lambda_{AF}\,\sigma_z .
\end{equation}

\begin{figure*}[ht!]
\centering
\includegraphics[width=0.45\linewidth]{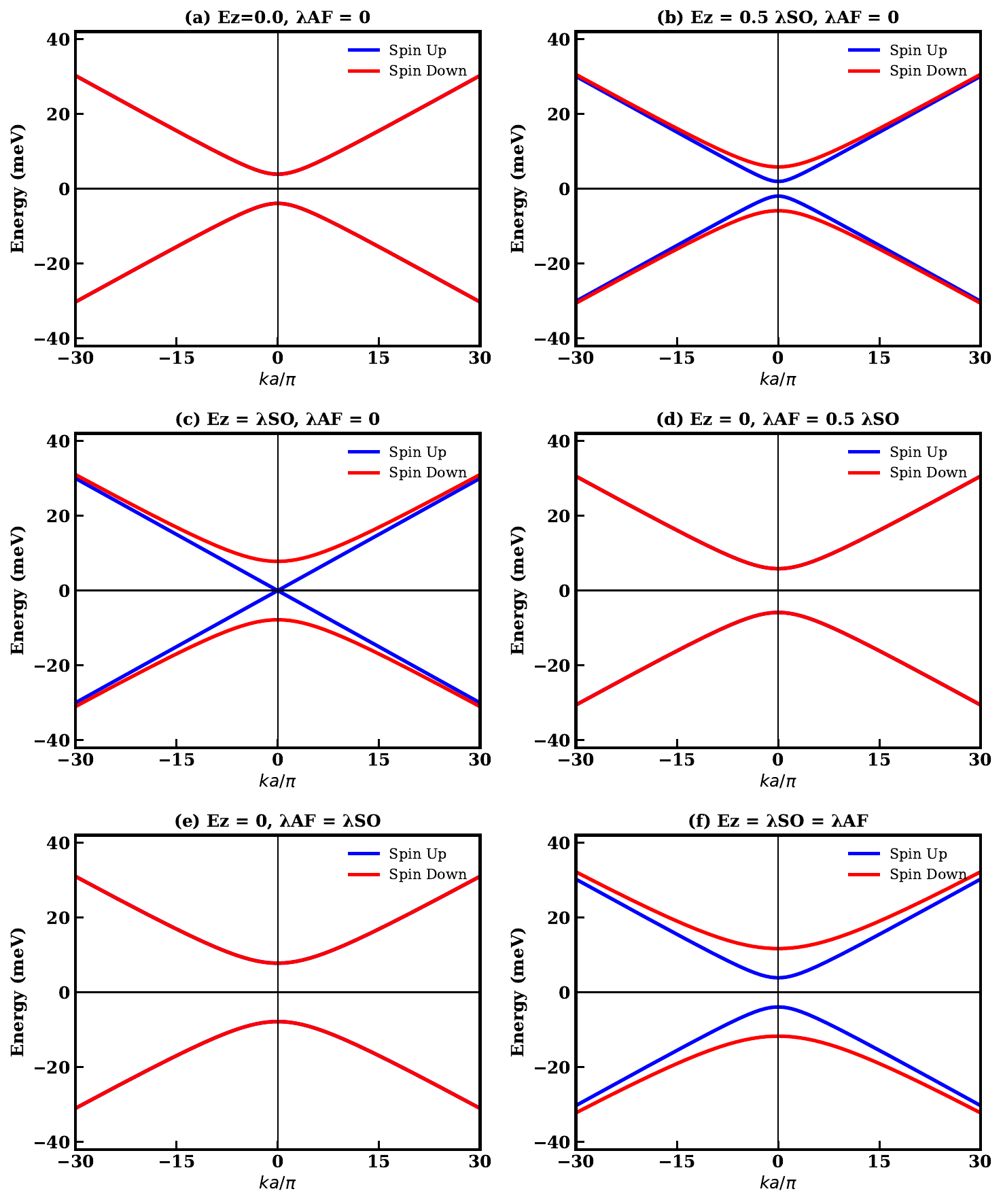}
\includegraphics[width=0.45\linewidth]{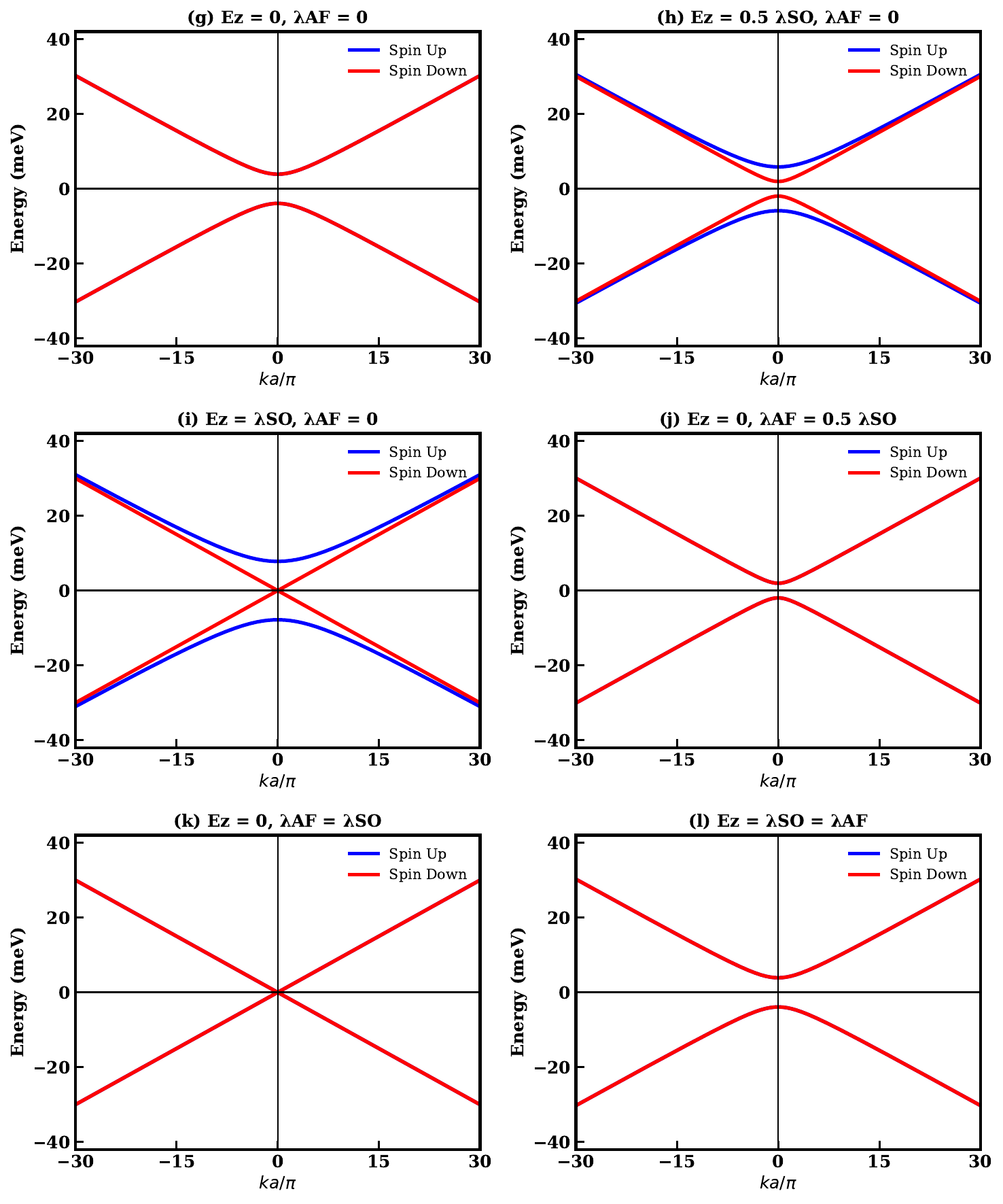}
\caption{Band structures near the $K$ and $K'$ valleys for distinct topological regimes. 
(a) $K$ valley SOC-dominated regime supporting the quantum spin Hall phase. 
(b) Reduced-gap QSH at $E_z\ell=0.5\lambda_{SO}$. 
(c) Critical point at $E_z\ell=\lambda_{SO}$ where the band gap closes.
(d) Quantum spin Hall regime with antiferromagnetic exchange. 
(e) AF insulator obtained by tuning $\lambda_{AF}=\lambda_{SO}$. 
(f) Anomalous Hall phase when both $E_z\ell$ and $\lambda_{AF}$ are present. 
(g) $K'$ valley SOC-dominated regime supporting the quantum spin Hall phase.
(h) Reduced-gap QSH at $E_z\ell=0.5\lambda_{SO}$. 
(i) Critical point at $E_z\ell=\lambda_{SO}$ where the band gap closes.
(j) Quantum spin Hall regime with antiferromagnetic exchange.
(k) Critical point obtained by tuning $\lambda_{AF}=\lambda_{SO}$.
(l) $K'$ valley Anomalous Hall phase when both $E_z\ell$ and $\lambda_{AF}$ are present.}
\label{fig:band_cases}
\end{figure*}

The first term is the usual Dirac kinetic part inherited from nearest-neighbour hopping:
$\eta=\pm1$ labels the two valleys ($K$ and $K'$), $v_F$ is the
Fermi velocity and being $\simeq 5.5\times10^5~\mathrm{m/s}$ in silicene, $\boldsymbol{\sigma}=(\sigma_x,\sigma_y,\sigma_z)$ are Pauli matrices
acting on the sublattice (pseudospin) degree of freedom, and $\mathbf{k}=(k_x,k_y)$ is the in-plane
momentum measured relative to the chosen Dirac point \cite{xu2020topology}.
The second term is the intrinsic SOC of
Kane--Mele type, with coupling $\lambda_{SO}=3.9~\mathrm{meV}$; the real-spin index is $s=\pm1$
for spin-up/spin-down \cite{PhysRevLett.107.076802}.
The third term captures the electric-field control: a perpendicular
field $E_z$ shifts the two sublattices differently because they sit at different heights, producing
a staggered potential of magnitude $E_z\ell$ (with $2\ell\simeq 0.46~\text{\AA}$ the vertical $A$--$B$
separation). Finally, the last term represents the antiferromagnetic exchange contribution, set by
$\lambda_{AF}$ \cite{ezawa2013spin} with the adjustable Dirac mass
\begin{equation}
    \Delta \eta_s = \eta s \lambda_{\mathrm{SO}} - E_z \ell  + s  \lambda_{\mathrm{AF}},
\end{equation}
where $\Delta \eta_s$ controls the gap in that spin-valley sector; the band gap closes when $\Delta \eta_s = 0$, which defines the phase boundaries in the $(E_z, \lambda_{\mathrm{AF}})$ parameter space. The resulting band energies are given by

\begin{equation}
\label{dispersion}
E_{\eta s} (k) = \pm \sqrt{(\hbar v_F k)^2 + \Delta \eta_s^2},
\end{equation}
where $k = \sqrt{k_x^2 + k_y^2}$
is the momentum magnitude in the Brillouin zone and $\pm$ represents the conduction (valence) band.

%====figg

For the sake of completeness, we first discuss the band structure of a buckled two-dimensional system for distinct topological regimes. Figures~\ref{fig:band_cases}(a)-(l) illustrate the effects of the two control parameters in various phases. A moderate $E_z\ell$ breaks inversion symmetry and generates unequal but finite gaps at the $K$ and $K'$ valleys. As $E_z\ell$ increases to a scale comparable to the spin–orbit coupling (SOC), the gap in one valley closes and subsequently reopens with an inverted sign, signaling an electric-field-driven topological transition \cite{ezawa2012topological}.

A similar evolution arises upon tuning $\lambda_{AF}$ at $E_z\ell = 0$; however, since $\lambda_{AF}$ breaks time-reversal symmetry, it redistributes the gaps asymmetrically between the valleys and can induce a valley-selective gap closing at $\lambda_{AF} = \lambda_{SO}$. This results in a gapless Dirac state in one valley, while the other remains gapped.

When both $E_z\ell$ and $\lambda_{AF}$ are present, the Dirac mass becomes a superposition of competing and reinforcing contributions, enabling regimes in which one valley is strongly gapped while the other remains only moderately gapped. Monitoring $|\Delta_{\eta s}|$ thus provides a transparent framework for predicting gap closings and reopenings, valley asymmetry, and the associated topological reorganization governing the Berry curvature and Hall responses.

The corresponding normalized choice of eigenstates is therefore
\begin{equation}
\ket{\psi^{+}_{\eta s}(\mathbf{k})}=
\begin{pmatrix}
\eta\,u_k\\[4pt]
v_k\,e^{+i\eta\phi_{\mathbf{k}}}
\end{pmatrix},
\qquad
\ket{\psi^{-}_{\eta s}(\mathbf{k})}=
\begin{pmatrix}
v_k\\[4pt]
-\eta\,u_k\,e^{+i\eta\phi_{\mathbf{k}}}
\end{pmatrix},
\end{equation}
with
\begin{equation}
u_k=\sqrt{\frac{1}{2}\left(1+\frac{\Delta_{\eta s}}{\varepsilon_{\eta s}(\mathbf{k})}\right)},
\qquad
v_k=\sqrt{\frac{1}{2}\left(1-\frac{\Delta_{\eta s}}{\varepsilon_{\eta s}(\mathbf{k})}\right)}.
\end{equation}
These spinors are defined up to an overall $\mathbf{k}$-dependent phase (gauge).

It is well known that Berry curvature plays a key role in transport when electrons are driven by a mechanical force \cite{xiao2006berry, xiao2007valley}. Because of this curvature, the group velocity of a Bloch electron gains an additional transverse component, causing the electron to deflect sideways during its motion. In simple terms, the Berry curvature describes how the phase of the electron’s wavefunction twists or changes in momentum space. It is defined as \cite{xiao2006berry, zhang2009berry, RevModPhys.82.1950}
\begin{equation}
\Omega(\mathbf{k}) = \nabla \times \langle u(\mathbf{k}) | i \nabla_{\mathbf{k}} | u(\mathbf{k}) \rangle.
\end{equation}
The spin and valley-dependent Berry curvature can be obtained analytically, which is written as
\begin{equation}
\Omega^{z}_{\eta,s} =\pm\eta
\frac{\hbar^{2} v_{F}^{2}\,\Delta_{\eta,s}}
{2\left[(\hbar v_{F} k)^{2} + \Delta_{\eta,s}^{2}\right]^{3/2}}.
\label{BC1}
\end{equation}
While the orbital magnetic moment in the semiclassical approach is expressed as~\cite{PhysRevB.107.235417}
\begin{equation}
m_{v,\eta s, z}(k) = \eta \frac{e \hbar v_F^{2} \Delta_{\eta s}}{2E_{\eta s}(k)^{2}}.
\label{OMM1}
\end{equation}
Having calculated the quantities in Eqs. (\ref{BC1}) and (\ref{OMM1}), the orbital Hall conductivity (OHC) can be easily obtained as $\sigma^{L_z}_{\eta s}=m_{v,\eta s, z}(k) \Omega^{z}_{\eta,s}(k)$ for a two-band model within the semiclassical approach \cite{bhowal2021orbital}. Notice that this expression overestimates the OHC; we will address the quantum correction to the OHC in the following discussion. By varying the physical parameters in the Hamiltonian, different phases emerge, including the quantum spin Hall (QSH), quantum valley Hall (QVH), antiferromagnetic quantum spin Hall (AF-QSH), quantum anomalous Hall (QAH), and critical phases. The Berry curvature and semiclassical orbital magnetic moment results for various phases are presented in the Appendix.
These results show that the Berry curvature is strongly enhanced as $|\Delta_{\eta s}|$ decreases, making the vicinity of the Dirac points the dominant contributor to the orbital Hall conductivity. Tuning $\lambda_{SO}$, $E_z$, and $\lambda_{AF}$ modifies the mass term and redistributes the Berry curvature among spin and valley sectors \cite{RevModPhys.82.1959,xiao2007valley}, providing an efficient means to control the orbital Hall response.

In the presence of an applied in-plane electric field, electrons acquire a transverse flow of orbital angular momentum, resulting in opposite orbital accumulation at the sample edges and the emergence of a transverse orbital current. In multiorbital systems, Bloch states exhibit momentum-dependent orbital character (orbital texture), which provides an intrinsic mechanism for generating orbital currents under an external field \cite{PhysRevLett.121.086602}. The intrinsic orbital Hall conductivity can therefore be expressed in a Berry-curvature-like form.

In the linear-response framework, the orbital Hall effect is defined as the transverse flow of orbital angular momentum induced by an electric field, $j^{L_z}_x \;=\; \sigma^{z,\mathrm{orb}}_{xy}\,E_y$
where the intrinsic orbital Hall conductivity is given by

\begin{equation}
\label{ohc gen}
\sigma^{z,\mathrm{orb}}_{xy}
=
-\frac{e}{(2\pi)^2}\sum_n\int_{\mathrm{BZ}} d^2k\;
f\!\left(\varepsilon_{n\bm k}-\mu\right)\,
\Omega^{z,\mathrm{orb}}_{n,xy}(\bm k).
\end{equation}
%============
where $f_{n\mathbf{k}}$ is the Fermi--Dirac occupation factor. From the Berry curvature, Eq. (\ref{BC1}), it is clear that $\sigma^{z,\mathrm{orb}}_{xy}\propto\Delta_{\eta s}^{-1}$. Thus, when the system approaches a topological phase transition, the Berry curvature becomes sharply peaked near the Dirac point, and the OHC is strongly enhanced. The central quantity entering the response is the orbital Berry curvature, $\Omega^{z,\mathrm{orb}}_{n,xy}(\bm k)$, obtained by replacing one velocity operator in the conventional Kubo formula with the orbital current operator \cite{shi2006proper}. It constitutes a geometric quantity defined in terms of band-structure matrix elements of the orbital angular momentum operator.
The orbital current operator is defined as
\begin{equation}
\hat{J}^{L_z}_y = \frac{1}{2} \left( \hat{v}_y \hat{L}_z + \hat{L}_z \hat{v}_y \right), 
\end{equation}
which ensures a Hermitian definition of the current.

%===========

The orbital Berry curvature is obtained from the Kubo formula by replacing one velocity operator with the orbital current operator \cite{bhowal2021orbital} ,

\begin{equation}
\label{kubo orb}
\Omega^{z,\mathrm{orb}}_{n,xy}(\bm {k})
=
2\hbar\,\mathrm{Im}\sum_{m\neq n}
\frac{
\langle n\bm k|\hat J^{z,\mathrm{orb}}_x|m\bm k\rangle\;
\langle m\bm k|\hat v_y|n\bm k\rangle
}{
\left(\varepsilon_{m\bm k}-\varepsilon_{n\bm k}\right)^2
}.
\end{equation}
%==================={=
The velocity operators are obtained from the Bloch Hamiltonian as $\hat{v}_x = \frac{\partial H}{\partial p_x} = \eta \sigma_x$, and $\hat{v}_y = \sigma_y$ where $p_i = \hbar v_F k_i$.

For Bloch electrons, the orbital angular momentum can be expressed in terms of the orbital magnetic moment~\cite{cysne2025description}. Two additional contributions are identified: for non-degenerate states, the first restores gauge covariance, while the second, also gauge covariant, can provide significant quantitative corrections depending on the system under consideration. Therefore, the matrix elements of the orbital magnetic moment are given by

\begin{equation}
m^{z}_{nn'}(\mathbf{k}) 
= -\frac{e}{4} \left\langle n \left| \left( \hat{\mathbf{r}} \times \hat{\mathbf{v}} \right) - \left( \hat{\mathbf{v}} \times \hat{\mathbf{r}} \right) \right| n' \right\rangle_z,
\end{equation}
This quantity can be decomposed into self-rotation and gauge-dependent contributions as

\begin{equation}
m_{nn'}(\mathbf{k}) =
m^{\mathrm{SR}}_{nn'}(\mathbf{k})
+ g^{\mathrm{I}}_{nn'}(\mathbf{k})
+ g^{\mathrm{II}}_{nn'}(\mathbf{k}).
\end{equation}
The first term, \(m^{\mathrm{SR}}_{nn'}(\mathbf{k})\), is the self rotation contribution obtained in the semiclassical wave-packet approach and is given by
\begin{equation}
m^{\mathrm{SR}}_{nn'}(\mathbf{k}) 
= -\frac{ie}{2\hbar}
\left\langle \nabla_{\mathbf{k}} n \left|
\times \left( \hat{H}_{\mathbf{k}} - \frac{\epsilon_n + \epsilon_{n'}}{2} \right)
\right| \nabla_{\mathbf{k}} n' \right\rangle,
\label{eq:msr}
\end{equation}
where $\nabla_{\mathbf{k}} = \hat{x}{\partial}/{\partial k_x} + \hat{y}{\partial}/{\partial k_y}$
denotes the ordinary momentum-space derivative, and \(\times\) represents the cross product. Here we are working within a fully quantum mechanical framework. Unlike the semiclassical approach, the non-zero matrix elements are not limited to a quasi-degenerate Hilbert subspace. As a result, and as highlighted in Ref. \cite{pozo2023multipole}, Eq. (\ref{eq:msr}) by itself does not satisfy gauge covariance. The second contribution is given by
\begin{equation}
g^{I}_{nn'}(\mathbf{k}) 
= \frac{e}{4} 
\left[ \left( \mathbf{A}_n + \mathbf{A}_{n'} \right) \times \mathbf{v}_{nn'} \right]_z 
(1 - \delta_{nn'}),
\label{eq:g1}
\end{equation}
where the Berry connection is defined as $\mathbf{A}_n = i \langle n | \nabla_{\mathbf{k}} n \rangle$ and the velocity matrix elements are $\mathbf{v}_{nn'} = \langle n | \hat{\mathbf{v}} | n' \rangle$.

As in \cite{pozo2023multipole}, the sum \(m^{\mathrm{SR}}_{nn'} + g^{I}_{nn'}\) forms a gauge-covariant quantity and takes a form similar to Eq. (\ref{eq:msr}) which can be written in terms of the covariant derivative $|D_{\mathbf{k}} n \rangle = |\nabla_{\mathbf{k}} n \rangle + i \mathbf{A}_n |n\rangle$.
In addition, there exists a second gauge-dependent contribution
\begin{equation}
g^{II}_{nn'}(\mathbf{k}) 
= \frac{e}{4}
\left[ \left( \mathbf{v}_{nn} + \mathbf{v}_{n'n'} \right) 
\times i \langle n | \nabla_{\mathbf{k}} n' \rangle \right]_z
(1 - \delta_{nn'}).
\label{eq:g2}
\end{equation}
Using this definition, the covariant form of the orbital magnetic moment is given by
\begin{equation}
m^{\mathrm{cov}}_{nn'}(\mathbf{k}) 
= -\frac{ie}{2\hbar}
\left\langle D_{\mathbf{k}} n\left|
\times \left( \hat{H}_{\mathbf{k}} - \frac{\epsilon_n + \epsilon_{n'}}{2} \right)
\right|D_{\mathbf{k}} n' \right\rangle.
\label{eq:mcov}
\end{equation}
Therefore, the full orbital magnetic moment can be written as
\begin{equation}
m_{nn'}(\mathbf{k}) = m^{\mathrm{cov}}_{nn'}(\mathbf{k}) + g^{II}_{nn'}(\mathbf{k}).
\label{eq:full_moment}
\end{equation}
Our numerical calculations show that $g^{I}_{nn'}(\mathbf{k})+g^{II}_{nn'}(\mathbf{k})$ reduces the \(m^{\mathrm{SR}}_{nn'}(\mathbf{k})\) by approximately $50\%$, indicating a significant contribution.  Moreover, the orbital angular momentum operator is related to the magnetic moment through
\begin{equation}
L^z_{nn'}(\mathbf{k}) =
-\frac{\hbar}{g_L \mu_B}
\, m_{nn'}(\mathbf{k}). 
\end{equation}

For numerical evaluation, the Hamiltonian is diagonalized at each momentum point, and the orbital Berry curvature is computed from interband matrix elements. In a two-band system, only transitions between the valence and conduction bands ($n, m \in {-, +}$) contribute, yielding
\begin{equation}
\Omega^{L_z}_{yx,-}
=
2\,\mathrm{Im}
\left[
\frac{
\langle u_- | \hat{J}^{L_z}_y | u_+ \rangle
\langle u_+ | \hat{v}_x | u_- \rangle
}{
\left( \epsilon_- - \epsilon_+  \right)^2
}
\right],
\end{equation}
and 
\begin{equation}
\Omega^{L_z}_{yx,+}
=
2\,\mathrm{Im}
\left[
\frac{
\langle u_+ | \hat{J}^{L_z}_y | u_- \rangle
\langle u_- | \hat{v}_x | u_+ \rangle
}{
\left( \epsilon_+ - \epsilon_-  \right)^2
}
\right].
\end{equation}
For continuum Dirac systems with rotational symmetry, the two-dimensional momentum integral reduces to a radial form, and accordingly, the OHC for a given spin and valley channel becomes
\begin{equation}
\sigma^{L_z}_{yx,\eta s}
=
e \int_0^{p_{\mathrm{max}}}
\frac{ p \, dp}{(2\pi)}
\left[
f(\epsilon_-)\,\Omega^{L_z}_{yx,-}(p)
+
f(\epsilon_+)\,\Omega^{L_z}_{yx,+}(p)
\right].
\end{equation}
The total conductivity is thus obtained by summing over all spin and valley indices,

\begin{equation}
\sigma^{L_z}_{yx,\mathrm{tot}}
=
\sum_{\eta = \pm 1}
\sum_{s = \pm 1}
\sigma^{L_z}_{yx,\eta s},
\end{equation}
and in our results, it is expressed in dimensionless units of $e/2\pi$. This expression indicates that the orbital Hall conductivity is determined by the orbital Berry curvature weighted by band occupations, with dominant contributions from states near the band edges. As the gap closes near topological phase transitions, the Berry curvature is strongly enhanced, resulting in a significant increase in conductivity. Tuning $E_z$ and $\lambda_{AF}$ thus offers an efficient route to control the magnitude and sign of the orbital Hall response across spin and valley sectors.
%============={{{{{{}}}}}}

\section{Numerical Results and Discussion}
We calculate the orbital Hall conductivity $\sigma_{xy}^{L_z}$ for a 2D buckled honeycomb lattice across five distinct electronic phases each realized by a different combination of vertically applied electric field $E_z$ and antiferromagnetic exchange strength $\lambda_{\text{AF}}$ with fixed spin–orbit coupling at $\lambda_{\mathrm{SO}}=3.9~\mathrm{meV}$. The OHC is evaluated as a function of Fermi energy $E_F$ and temperature $T$ for each case, revealing how band structure, physical quantities, and symmetry-breaking fields influence orbital transport. At low temperature and finite Fermi energy, we use an approximated chemical potential expression as $\mu(T)=E_F-\pi^2 (k_B T)^2/6E_F$ \cite{ramezanali2009finite}.

We list the system's topological features in Table I to offer a more thorough categorization of the various electrical regimes. The behavior of the Dirac mass term ( $\Delta_{\eta,s}$ ), the existence or lack of symmetry (inversion and time-reversal), and the associated topological invariants are used to identify each phase. Specifically, the QVH phase has a finite valley Chern number due to inversion symmetry breaking, whereas the QSH phase has a nontrivial spin Chern number despite a vanishing total Chern number. The system can enter a QAH phase with a finite total Chern number when the antiferromagnetic exchange field breaks time-reversal symmetry. The system becomes gapless, and the topological invariants are ill-defined at critical points where the Dirac mass disappears in one or more spin-valley sectors.

\begin{widetext}
    
\begin{table}[t]
\centering
\caption{Topological classification of electronic phases in the buckled 2D Dirac system. The phases are characterized by the Dirac mass term $\Delta_{\eta,s}$, symmetry properties, and associated topological invariants.}
\begin{tabular}{lccccccc}
\hline\hline
Phase & Condition & Gap & TRS & IS & $C$ & $C_s$ & $C_v$ \\
\hline
QSH 
& $|E_z \ell| < \lambda_{SO},\ \lambda_{AF}=0$ 
& Finite 
& Yes 
& Yes 
& 0 
& $\neq 0$ 
& 0 \\

QVH 
& $|E_z \ell| > \lambda_{SO},\ \lambda_{AF}=0$ 
& Finite 
& Yes 
& No 
& 0 
& 0 
& $\neq 0$ \\

AF-QSH-like 
& $\lambda_{AF} < \lambda_{SO},\ E_z=0$ 
& Finite 
& No 
& Yes 
& 0 
& $\neq 0$ 
& 0 \\

QAH 
& $\lambda_{AF} \sim \lambda_{SO},\ E_z \neq 0$ 
& Finite 
& No 
& No 
& $\neq 0$ 
& -- 
& -- \\

VP-QAH 
& $\lambda_{SO} \sim \ E_z \sim \lambda_{AF}\neq0$ 
& Finite 
& No 
& No 
& $\neq 0$
& $\neq 0$ 
& $\neq 0$\\

Critical 
& $\Delta_{\eta,s}=0$ 
& Zero 
& Depends 
& Depends 
& Undefined 
& Undefined 
& Undefined \\
\hline\hline
\end{tabular}
\label{tab:phases}
\end{table}
\end{widetext}

%=====figOHC
\begin{figure*}[ht!]
\centering
\includegraphics[width=\linewidth]{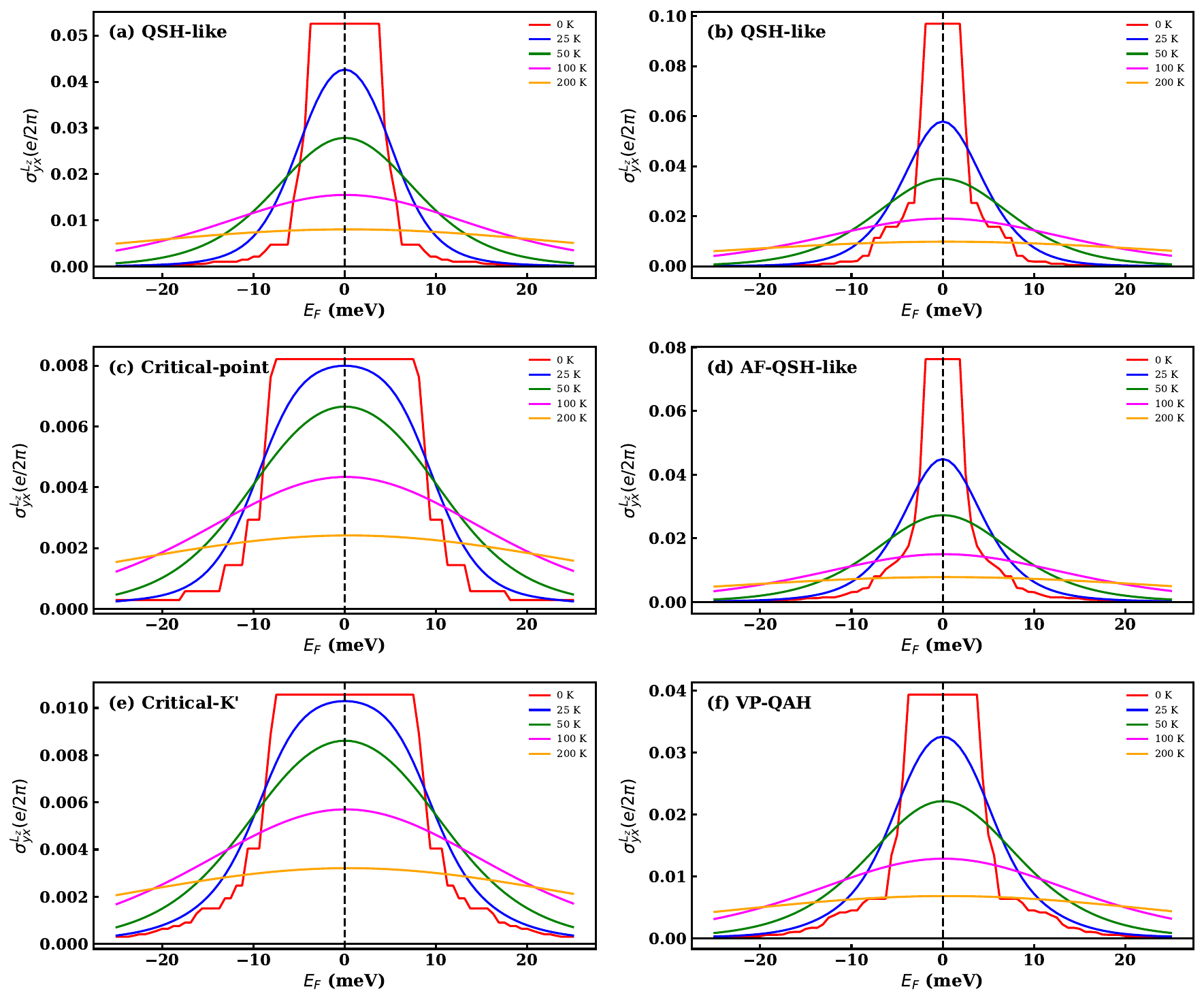}
\caption{Orbital Hall conductivity $\sigma_{yx}^{L_z}$ (in units of $e/2\pi$) as a function of $E_F$ for distinct phases at different temperatures. Notice that we use $T=0$, for the case that $T\rightarrow 0$ around $E_F=0$. 
    (a) SOC-dominated regime supporting the quantum spin Hall phase. 
    (b) Reduced-gap QSH at $E_z\ell=0.5\lambda_{SO}$. 
    (c) Critical point at $E_z\ell=\lambda_{SO}$.
    (d) Quantum spin Hall regime with antiferromagnetic exchange. 
    (e) Critical point at $\lambda_{AF}=\lambda_{SO}$. 
    (f) Valley Polarized Anomalous Hall phase when both $E_z\ell$ and $\lambda_{AF}$ are present. }
    \label{fig:ohc_cases}
\end{figure*}
\begin{figure*}[ht!]
\centering
\includegraphics[width=\linewidth]{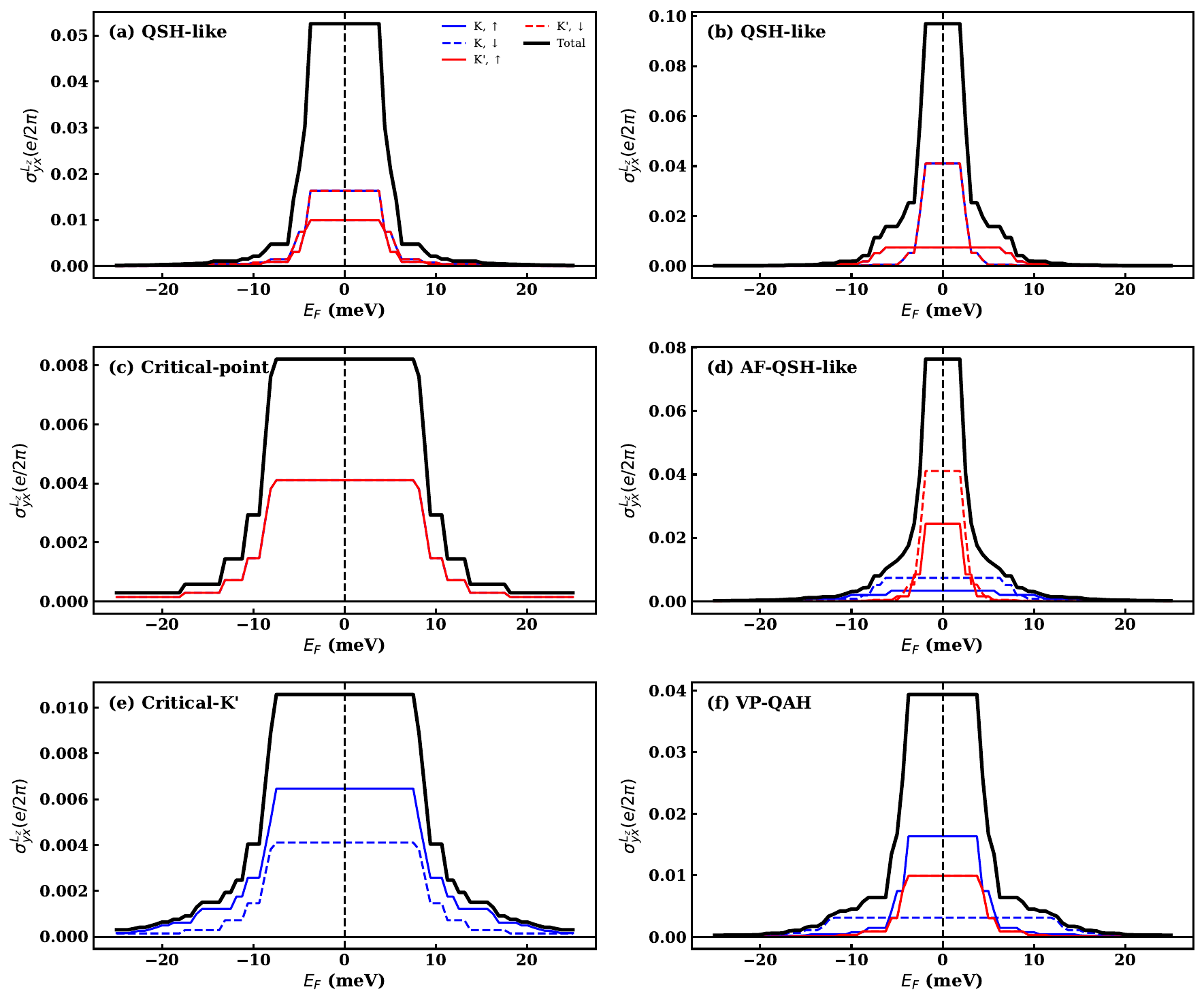}
\caption{Orbital Hall conductivity $\sigma_{yx}^{L_z}$ (in units of $e/2\pi$) as a function of $E_F$ near the $K$ and $K'$ valleys for distinct topological regimes. 
(a) SOC-dominated QSH phase. 
(b) Reduced-gap QSH at $E_z\ell=0.5\lambda_{SO}$. 
(c) Critical point at $E_z\ell=\lambda_{SO}$. 
(d) QSH with antiferromagnetic exchange. 
(e) AF insulator at $\lambda_{AF}=\lambda_{SO}$. 
(f) Valley polarized Anomalous Hall phase.}
\label{fig:OHC_valley}
\end{figure*}

In Fig. \ref{fig:ohc_cases}(a), where $E_z=0$ and $\lambda_{AF}=0$, the system is governed solely by the intrinsic spin–orbit coupling $\lambda_{SO}$. The band gap remains finite, and the system resides in the quantum QSH phase. In this regime, time-reversal symmetry is preserved, and the two spin channels contribute oppositely to the transverse response, resulting in a symmetric orbital Hall conductivity profile centered around $E_F=0$. The plateau-like structure observed at low temperature indicates a well-defined topological gap. At low temperatures, one can find that the leading correction behaves as
\begin{equation}
\sigma^{L_z}_{\eta s}(T) \approx \sigma^{L_z}_{\eta s}(0)
\left[ 1 - \alpha \left( \frac{k_B T}{|\Delta_{\eta s}|} \right)^2 \right],
\end{equation}
where $\alpha$ is a dimensionless constant of order unity that depends weakly on the cutoff and band details.

In Fig. \ref{fig:ohc_cases} (b), the OHC is presented for $\lambda_{\mathrm{AF}}=0$ and $E_z=\lambda_{\mathrm{SO}}/2$, where inversion symmetry is broken but time-reversal symmetry remains intact. When a moderate electric field is applied, the sublattice potential difference induced by the buckled structure modifies the Dirac mass term. Although the band gap decreases compared to the zero-field case, it remains open since $E_z < \lambda_{SO}$. Consequently, the system still lies in the QSH phase. In this regime, a strong OHC emerges due to the finite orbital Berry curvature and reaches its maximum when the Fermi level lies within the band gap \cite{bhowal2021orbital}. At very low temperature, $\sigma_{xy}^{L_z}$ shows a sharp or nearly flat peak around $E_F=0$, reflecting the large Berry curvature concentrated near the band edges. As the temperature increases and reduces the Fermi distribution contribution, this peak gradually broadens and decreases because thermal excitations smear the occupation of states, although the overall response remains clearly visible.

In contrast, the trivial insulating phase shown in Fig.~\ref{fig:ohc_cases}(c) exhibits a much weaker OHC. At this point, the electric field compensates the intrinsic spin–orbit coupling for one spin channel, causing the band gap to close at a valley point. This marks a topological phase transition from the QSH phase toward the QVH phase. While the overall shape is still centered around the gap, only a small plateau appears at $E_F=0$ at low temperature. This residual signal originates from valley-dependent Berry curvature caused by inversion symmetry breaking, but it quickly fades with increasing temperature and becomes almost negligible at room temperature.
When an antiferromagnetic exchange field is introduced without an electric field (Fig. \ref{fig:ohc_cases} (d), time-reversal symmetry is broken, yet the system retains an effective inversion symmetry. For $\lambda_{\mathrm{AF}}=\lambda_{\mathrm{SO}}/2$, the OHC remains large and behaves similarly to the topological case, showing that the key requirement is not symmetry itself but the presence of strong Berry curvature near the band edges.

In Fig.~\ref{fig:ohc_cases}(e), the exchange field is increased to $\lambda_{\mathrm{AF}}=\lambda_{\mathrm{SO}}$ while $E_z=0$. Under this condition, the gap associated with one spin channel closes, indicating a critical point separating distinct topological regimes. The conductivity plateau becomes less pronounced, and the magnitude of the orbital Hall response is significantly reduced, reflecting the transition between the QSH and QAH phases. At low temperature, only a weak residual signal remains due to partial cancellation between contributions from different spin channels, and this response is further suppressed as temperature increases.

Finally, in the mixed regime shown in Fig.~\ref{fig:ohc_cases}(f), where $E_z=\lambda_{\mathrm{SO}}$ and $\lambda_{\mathrm{AF}}=\lambda_{\mathrm{SO}}$, the combined electric and exchange fields break time-reversal symmetry and lift both spin and valley degeneracies, driving the system into the quantum anomalous Hall phase. In this regime, the orbital Hall conductivity remains finite but exhibits a moderate peak structure arising from the asymmetric band topology. As in the other cases, increasing temperature gradually suppresses the response due to thermal broadening around the band edges.

Figures~\ref{fig:OHC_valley}(a)-(f) show the orbital Hall conductivity $\sigma_{yx}^{L_{z}}$ as a function of the Fermi energy $E_{F}$ for different topological phases using the same set of parameters. In contrast to the previous figure, where temperature dependence was emphasized, here we focus on the individual valley–spin resolved contributions to the orbital Hall response. The colored curves represent the four valley–spin channels ($K\uparrow$, $K\downarrow$, $K'\uparrow$, $K'\downarrow$), while the black curve denotes the total orbital Hall conductivity obtained by summing all channels.

In the QSH-like regimes shown in Fig.~\ref{fig:OHC_valley}(a) and (b), the valley–spin contributions are nearly symmetric, resulting in a robust plateau of the total conductivity around the charge neutrality point, which originates from the Berry curvature of the gapped Dirac bands. At the critical point in Fig.~\ref{fig:OHC_valley}(c), the orbital Hall response is reduced and becomes more symmetric as the band gap approaches closure. When antiferromagnetic exchange is introduced, as shown in Fig.~\ref{fig:OHC_valley}(d), the spin–valley degeneracy is lifted, leading to an asymmetric distribution of contributions in which certain channels ($K\downarrow,K\uparrow$) dominate the orbital transport.

In the valley-selective critical regime shown in Fig.~\ref{fig:OHC_valley}(e), the orbital response is predominantly governed by a single valley channel, while the others become strongly suppressed. Finally, in the mixed anomalous quantum Hall phase in Fig.~\ref{fig:OHC_valley}(f), multiple valley–spin channels contribute with different magnitudes, giving rise to a finite total orbital Hall conductivity. Overall, these results highlight that the orbital Hall response is governed by the interplay between valley, spin, and the underlying topological phase.

\begin{figure*}[ht!]
\centering
\includegraphics[width=0.43\linewidth]{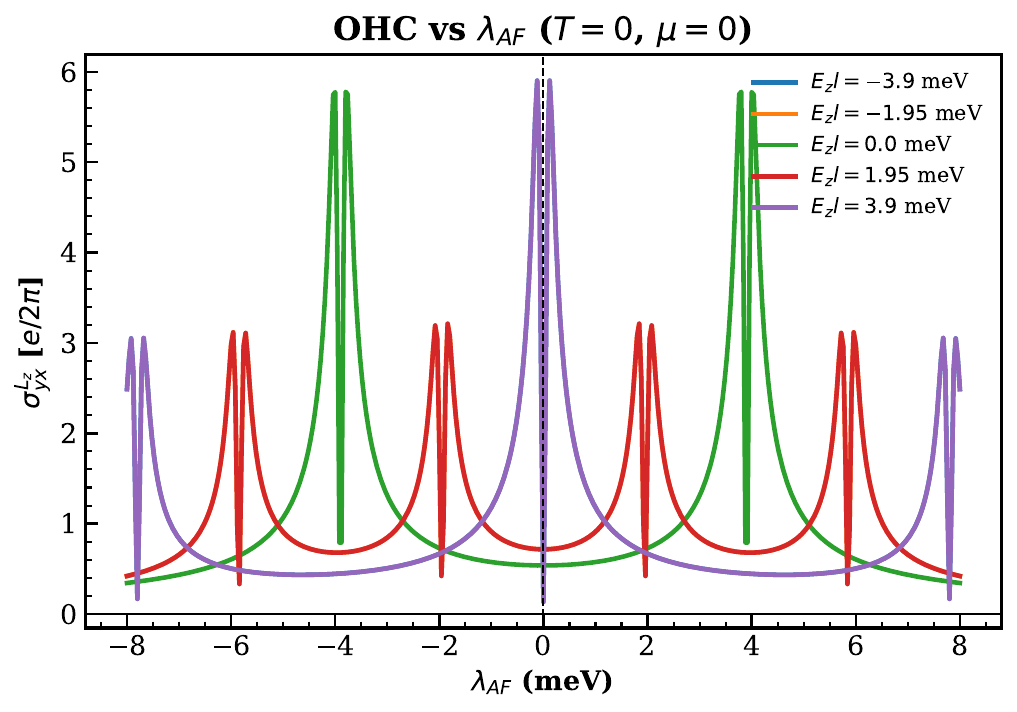}
\includegraphics[width=0.43\linewidth]{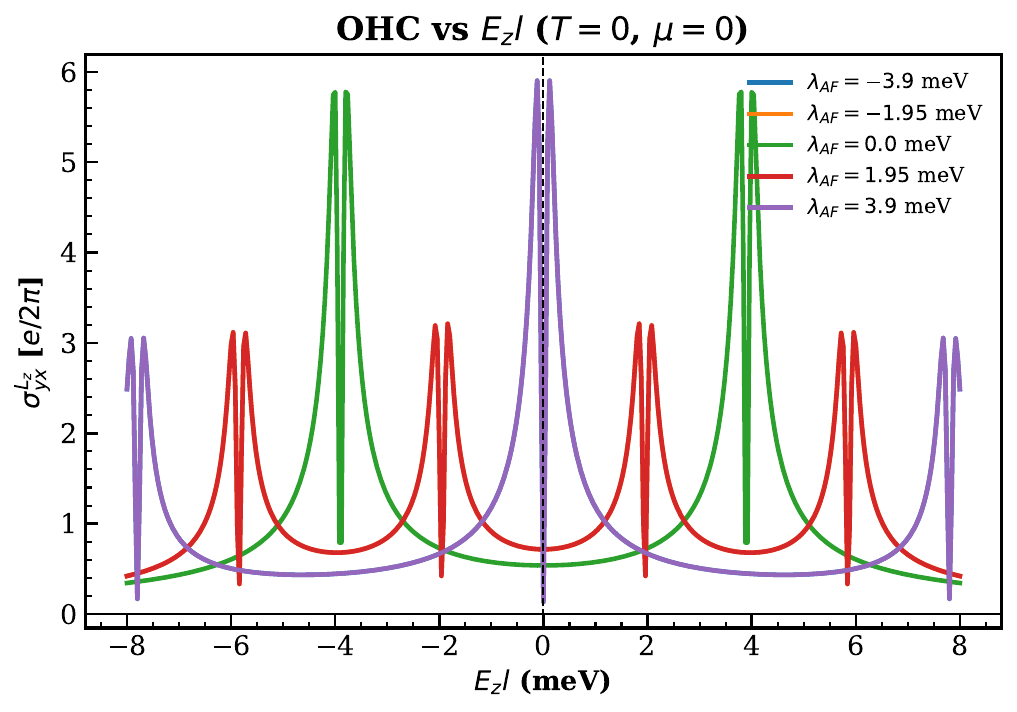}
\caption{Figure (a) shows Orbital Hall conductivity $\sigma_{yx}^{L_z}$ (in units of $e/2\pi$) as a function of $\lambda_{AF}$ for different values of Electric Field and (b) shows Orbital Hall conductivity $\sigma_{yx}^{L_z}$ (in units of $e/2\pi$) vs $E_z$ for different values of Exchange field.}
\label{fig:external_stimuli}
\end{figure*}

Figure~\ref{fig:external_stimuli} illustrates the orbital Hall conductivity at zero temperature as a function of the antiferromagnetic exchange coupling $\lambda_{\mathrm{AF}}$ and the perpendicular electric field $E_z$, evaluated at very small Fermi energy. In both panels, the conductivity exhibits sharp, symmetric peaks at specific values of the tuning parameters. These peaks occur when the band gap becomes very small or closes, leading to a strong enhancement and redistribution of Berry curvature, which in turn produces a large orbital Hall response. Since the orbital Hall conductivity in the present gapped Dirac system primarily originates from the intrinsic Berry curvature mechanism, the intrinsic contribution is expected to dominate in the insulating regime where Fermi surface effects are suppressed~\cite{burgos2024orbital}. Previous studies have shown that impurity scattering mainly leads to quantitative and not qualitative modifications of the orbital Hall response depending on the orbital texture and impurity scattering potential ~\cite{burgos2024orbital,tang2024role}. In the present work, we treat the electrons as independent and neglect explicit electron--electron interaction effects. A detailed investigation of these effects is beyond the scope of the present work.

In Fig.~\ref{fig:external_stimuli}(a), the peak positions shift as the electric field is varied, whereas in Fig.~\ref{fig:external_stimuli}(b) they move with the exchange coupling. This clearly demonstrates the interplay between $E_z$ and $\lambda_{\mathrm{AF}}$ in controlling the band structure and associated topological response. The symmetry of the curves reflects the underlying symmetries of the system, while the sharp features emphasize the absence of thermal broadening at zero temperature.

Overall, these results show that the orbital Hall effect can be effectively tuned by external fields and is closely tied to the evolution of Berry curvature near band crossings, consistent with earlier studies on orbital transport in Dirac and topological materials \cite{RevModPhys.82.1950,tanaka2008intrinsic,PhysRevLett.121.086602}.

The results clearly show that the intrinsic OHE in buckled honeycomb Dirac materials is mainly determined by band topology and symmetry. By tuning $E_z\ell$ and $\lambda_{\mathrm{AF}}$, the OHC changes systematically across different phases and becomes strongest when the system enters a band-inverted, topological regime. In particular, the quantum spin Hall phases display a much larger OHC, which can be understood as a direct consequence of Berry-curvature “hot spots” forming near the inverted band edges \cite{RevModPhys.82.1959}.

This behavior reflects how sensitive the OHC is to the orbital nature of the electronic states. In topological phases, Berry curvature is sharply concentrated near the $K$ and $K'$ valleys with opposite signs \cite{xiao2007valley}, and the contributions from different sectors tend to add up, giving a strong net response. In contrast, trivial phases lack band inversion, so the Berry curvature is weaker and more spread out, leading to significant cancellation between spin and valley contributions and, therefore, a much smaller OHC \cite{xiao2007valley,RevModPhys.82.1959}.

\begin{figure}[t!]
    \centering
    \includegraphics[width=0.50\textwidth]{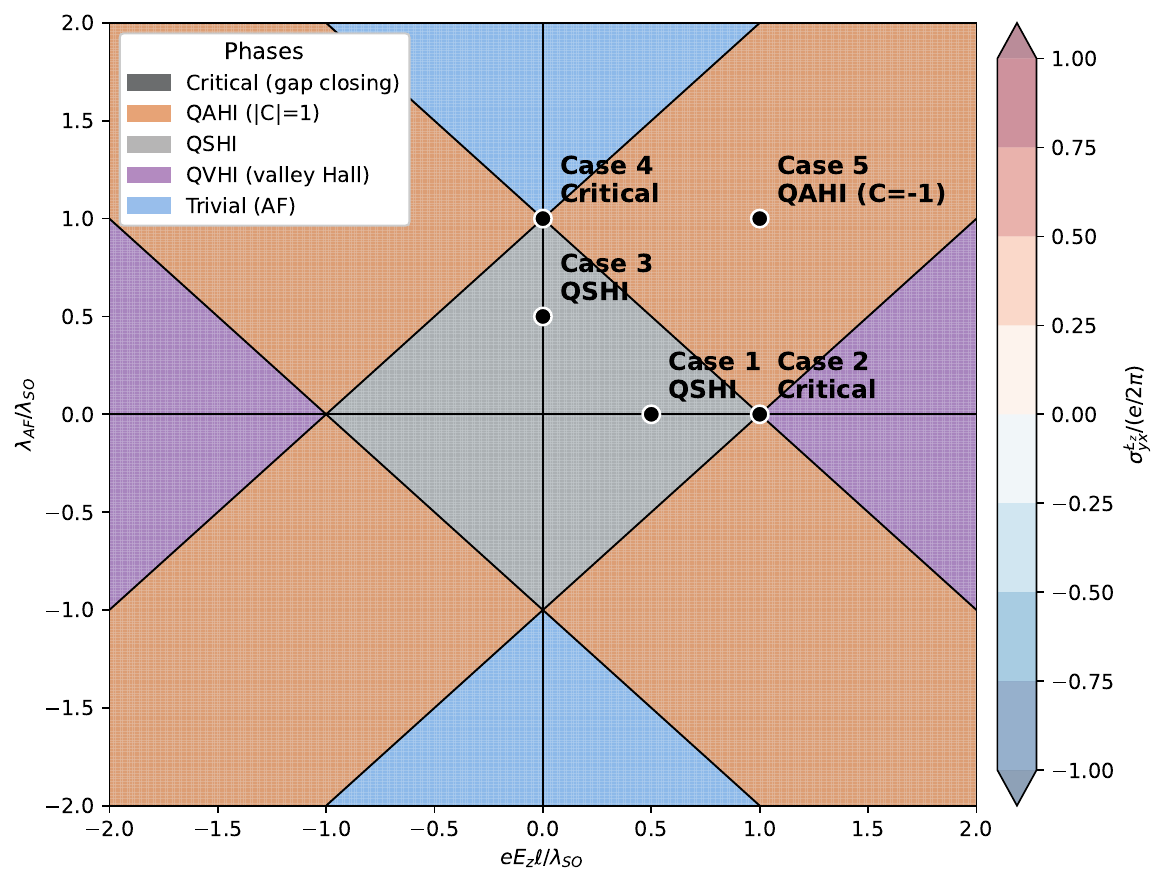}
    \caption{Phase diagram of the 2D buckled system as a function of the perpendicular electric field $E_z$ and the antiferromagnetic exchange field $\lambda_{\mathrm{AF}}$, showing QSH, QVHI, QAHI, and critical (gap-closing) phases. The corresponding heatmap of the orbital Hall conductivity $\sigma_{yx}^{L_z}$ (in units of $e/2\pi$) at the Fermi level is overlaid, illustrating its variation across different topological regimes.}
    \label{fig:phase_map}
\end{figure}

Close to phase transitions, where the gap closes and reopens, the Berry curvature is reshaped, causing rapid changes in $\sigma_{yx}^{L_z}$. Even though finite temperature smooths these features, the variation near these critical points remains a clear signature of topological switching \cite{RevModPhys.82.1959,ezawa2012spin}. Another important point is that inversion symmetry breaking is essential for obtaining a finite OHC, as it prevents complete cancellation of Berry curvature across the Brillouin zone. Time-reversal symmetry breaking, on the other hand, mainly redistributes contributions between different sectors and does not by itself control the strength of the response \cite{RevModPhys.82.1959}. Temperature tends to broaden and reduce $\sigma_{yx}^{L_z}$, but a significant part of the OHC survives even at room temperature in topological phases. This is because the Berry curvature remains strongly localized near the band edges, maintaining a net orbital response despite thermal excitations. In materials with larger gaps, this robustness is expected to be even more pronounced \citep{yao2026orbital}.

Overall, these results show that the intrinsic OHC is fundamentally linked to the Berry curvature of bulk bands and is maximized in band-inverted phases. Its strong dependence on $E_z\ell$ and $\lambda_{\mathrm{AF}}$ also means that orbital transport can be effectively tuned, making buckled 2D materials promising candidates for orbitronic applications, where orbital currents could be detected through spin conversion or magneto-optical methods \cite{PhysRevLett.121.086602,choi2023observation,lyalin2023magneto}.

In this work, we explore how a perpendicular electric field $E_z$ and an antiferromagnetic exchange field $\lambda_{\mathrm{AF}}$ together influence the orbital Hall conductivity in a 2D buckled material. The phase diagram in Fig.~\ref{fig:phase_map} highlights five distinct electronic phases, including QSH, QVHI, critical gap-closing states, and two QAHI phases. It provides a clear picture of how tuning these external parameters drives the system from one phase to another.

Alongside this, the heatmap of $\sigma_{yx}^{L_z}$ at the Fermi level shows how the orbital Hall response changes across the phase diagram. The conductivity does not vary randomly; instead, it follows the underlying changes in symmetry and band structure caused by $E_z$ and $\lambda_{\mathrm{AF}}$. In particular, the OHC becomes much stronger in topologically nontrivial phases such as QSH and QAHI, where the electronic structure supports large Berry curvature and robust transport features.

\section{Conclusion}
The intrinsic orbital Hall effect in a buckled two-dimensional Dirac system under the combined effects of an antiferromagnetic exchange field and a perpendicular electric field has been studied in this work. We have shown that the orbital Hall conductivity is very sensitive to the underlying band structure and can be efficiently controlled by externally controllable parameters within a low-energy massive Dirac framework and Berry-curvature-based linear response theory.

According to our findings, the orbital Hall response is significantly boosted in regimes where the Dirac mass decreases, especially near topological phase transitions and band inversion. We demonstrated distinct electronic phases, such as quantum spin Hall, quantum valley Hall, and quantum anomalous Hall regimes, by methodically investigating the parameter space defined by the electric and exchange fields. We have also established unambiguous correlations between these phases and distinctive characteristics of the orbital Hall conductivity. Specifically, we find that trivial insulating regimes show significantly reduced conductivity due to partial cancellation of contributions from different spin and valley sectors, whereas topologically nontrivial phases show prominent orbital responses associated with sharply localized Berry curvature near the Dirac points.

We also investigated the influence of temperature on OHC and showed that the qualitative characteristics associated with various phases remain robust over a broad temperature range, despite thermal broadening reducing the orbital Hall response's size and sharpness. The orbital Hall effect in buckled Dirac materials may be experimentally accessible under realistic settings, according to this robustness.

The current findings demonstrate that the orbital Hall conductivity is a sensitive indicator of symmetry breaking and band topology, even in a simple two-band model. However, our analysis is predicated on several simplifying assumptions, such as the intrinsic (disorder-free) limit and the disregard of extra spin-orbit interactions like Rashba coupling and electron-electron interactions. Future research should focus on incorporating these effects, material-specific modeling, and realistic device designs.

Our results give a conceptual foundation for connecting orbital Hall phenomena to programmable topological band characteristics and establish buckled two-dimensional materials as a flexible platform for building tunable orbital transport. The design of orbitronic functions in low-dimensional quantum materials, where orbital degrees of freedom can be used in addition to or instead of charge and spin, is made possible by this link.

%In this work, we explored how the orbital Hall effect evolves in buckled two-dimensional Dirac materials when external electric and antiferromagnetic exchange fields are introduced. Our results show that the orbital Hall conductivity is closely linked to the underlying band topology, with strong features appearing near the band edges due to Berry curvature effects. By tuning the external fields, the system can be driven through different topological regimes, including the quantum spin Hall phase, critical transition points, and the quantum anomalous Hall phase. An important outcome of this study is that the orbital Hall response itself clearly reflects these phase transitions, making it a useful probe of the system’s topological nature. We also find that increasing temperature smooths out the conductivity features, although the key topological behavior remains intact.

%Overall, these results highlight how effectively orbital transport can be controlled in buckled two-dimensional materials. This tunability makes such systems promising for future orbitronic applications, where orbital currents could be used for information processing. Looking ahead, it would be interesting to include additional effects such as Rashba spin–orbit coupling, disorder, and interactions to better connect with realistic systems. Extending this work to specific materials and experimental setups, as well as exploring external control methods like optical driving or strain, could further open up new possibilities for manipulating orbital degrees of freedom in next-generation devices.

\begin{figure*}[ht!]
\centering
\includegraphics[width=1.0\linewidth]{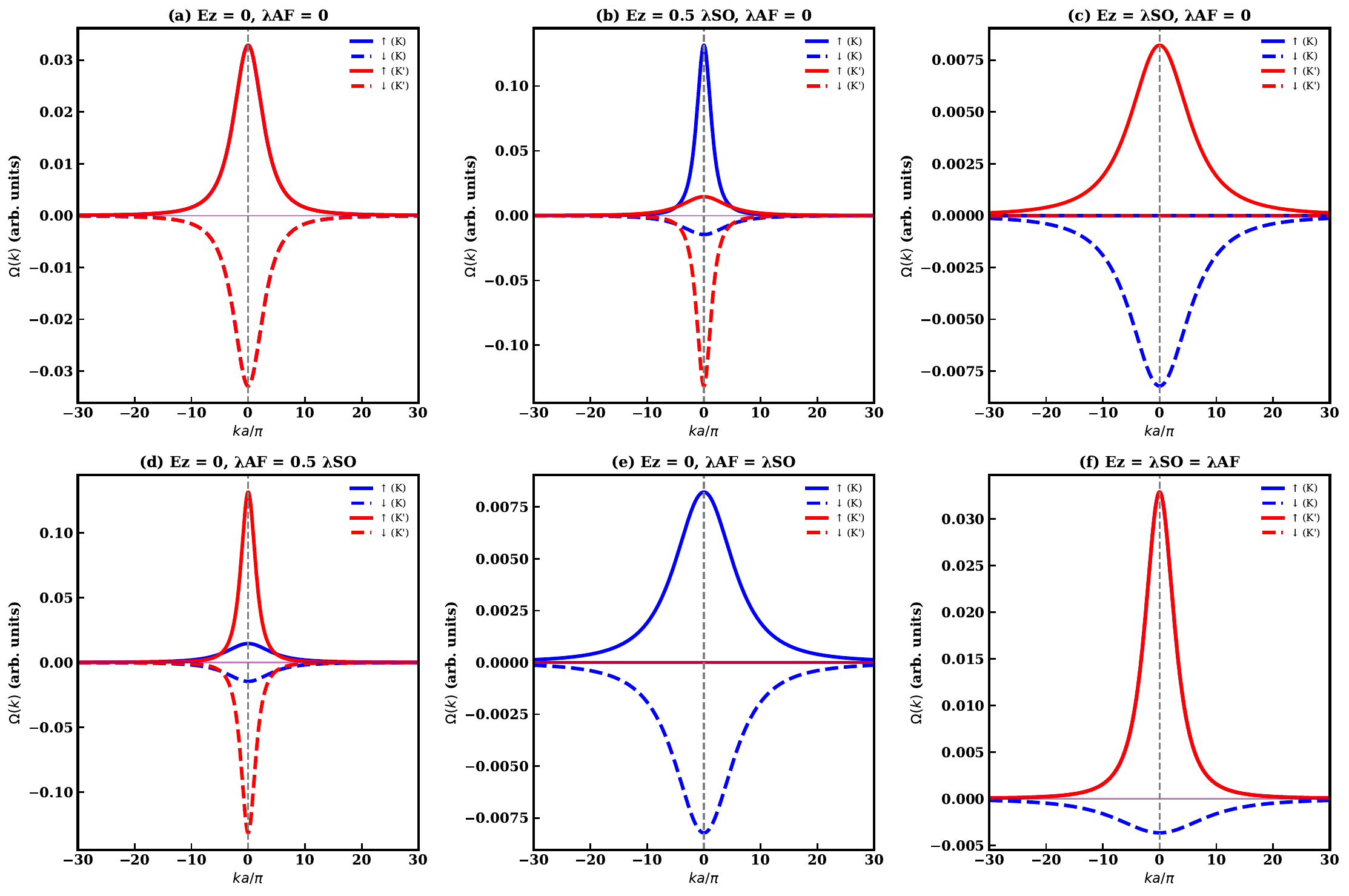}

\caption{Berry curvature $\Omega_z(k)$ for spin up/down in the $K$ and $K'$ valleys for five parameter sets. The curvature is sharply peaked near $k=0$ and is strongest when the effective mass $|\Delta_{\eta s}|$ is smallest. 
(a) SOC-dominated regime supporting the quantum spin Hall phase. 
(b) Reduced-gap QSH at $E_z\ell=0.5\lambda_{SO}$. 
(c) Critical point at $E_z\ell=\lambda_{SO}$ where the band gap closes.
(d) Quantum spin Hall regime with antiferromagnetic exchange. 
(e) Critical Point at $\lambda_{AF}=\lambda_{SO}$. 
(f) Anomalous Hall phase when both $E_z\ell$ and $\lambda_{AF}$ are present. 
}
\label{fig:berry_cases}
\end{figure*}
\newpage
\appendix
\section*{Appendix}
Here, we provide more information regarding the Berry curvature, Eq. (\ref{BC1}), and orbital magnetic moment within the semiclassical approach given by Eq. (\ref{OMM1}).

\begin{figure*}[ht!]
\centering
\includegraphics[width=1.0\linewidth]{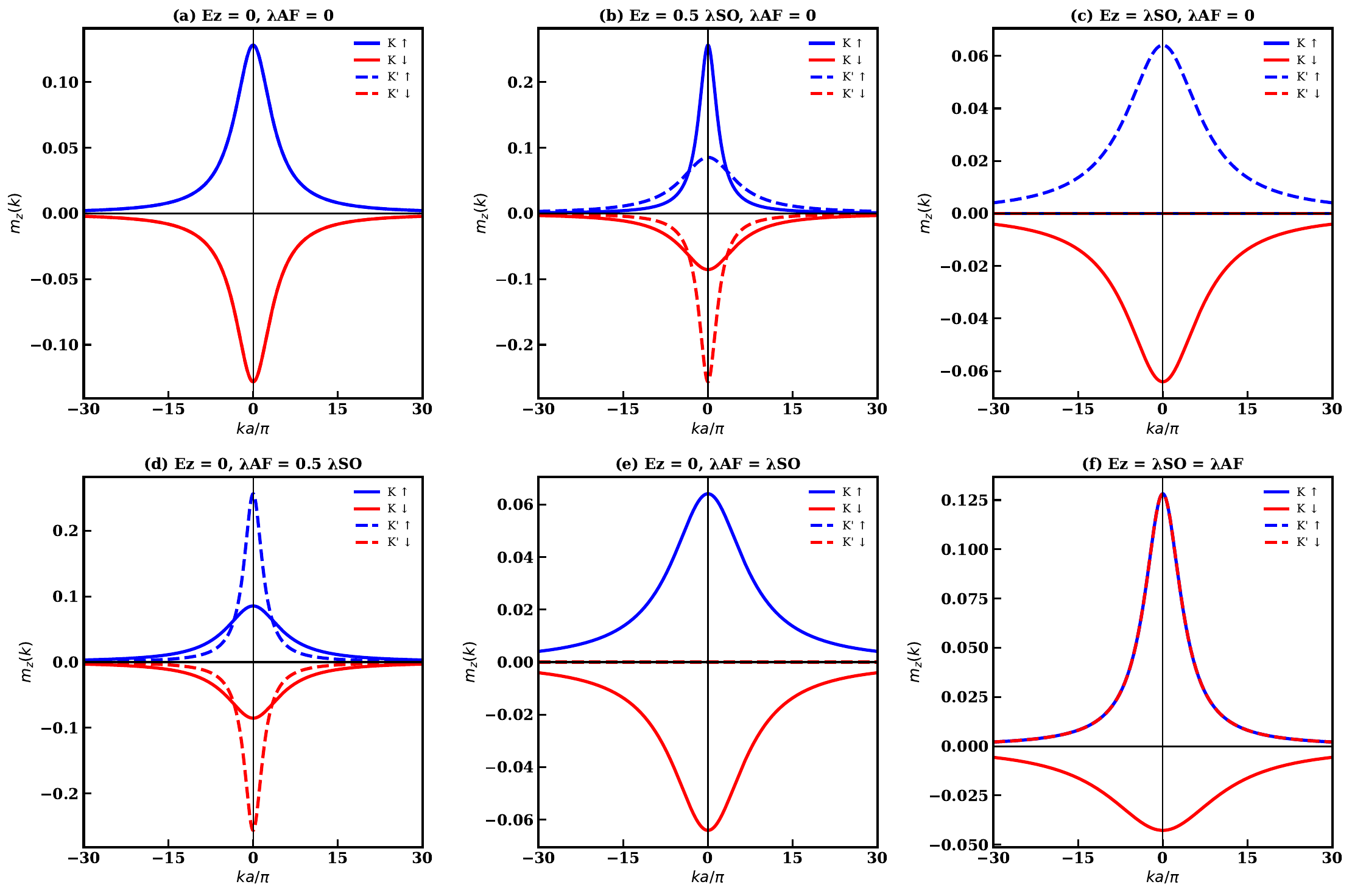}

\caption{Orbital magnetic moment $m_z(k)$ for spin up/down in the $K$ and $K'$ valleys for different topological regimes. (a) SOC-dominated gapped regime associated with the quantum spin Hall side in silicene-family. (b) Reduced-gap ASH. (c) $E_z\ell=\lambda_{SO}$ is a critical point. (d) QSH phase. (e) Critical point. (f) Anomalous Hall phase. }
\label{fig:magnetic}
\end{figure*}

In Fig. \ref{fig:berry_cases}(a), only spin-orbit coupling is present, and no external stimuli are applied to SOC dominating regime supporting the Quantum spin Hall Effect. In Fig. \ref{fig:berry_cases}(b), a moderate perpendicular electric field is applied while the antiferromagnetic (AF) exchange is set to zero ($\lambda_{\mathrm{AF}}=0$). The blue solid curve ($K$, spin up) shows a large positive peak because $|\Delta|$ is small ($\approx 1.95$~meV), so the curvature is strong near the Dirac point. The red dashed curve ($K'$, spin down) shows a large negative dip for the same reason: $|\Delta|$ is also small ($\approx 1.95$~meV), but the sign differs due to the valley factor, which matches the result of \cite{xu2020topology}. The remaining two curves (blue dashed and red solid) stay weak because their $|\Delta|$ is large ($\approx 5.85$~meV), meaning those sectors are more strongly gapped and therefore contribute less Berry curvature. This parameter set satisfies $|E_{z}\ell|<\lambda_{SO}$ (with $\lambda_{AF}=0$), meaning the system stays in the SOC-dominated gapped regime (often associated with the quantum spin Hall side in silicene-family models).

%..  

Figure \ref{fig:berry_cases}(c) shows that two sectors have $\Delta=0$, meaning the gap closes in those channels. For that reason, their Berry curvature
curves stay almost on the zero line, that is, the solid blue trace ($K\uparrow$) and the dashed red trace ($K'\downarrow$). The other two sectors
have a larger gap ($|\Delta|=7.8$~meV), so they still show a small Berry curvature near $k=0$. That is why the blue dashed curve
($K\downarrow$) shows a small negative dip, and the red solid curve ($K'\uparrow$) shows a small positive peak. Due to $E_z\ell$ inversion symmetries are broken here. The $K$ and $K'$ valleys typically contribute Berry curvature with opposite signs, so the curve in $K$ is negative while the corresponding curve in $K'$ is positive \cite{RevModPhys.82.1959,xiao2007valley}. From the topological point of view, $E_z\ell=\lambda_{SO}$ is a critical point because the band gap closes in two spin--valley channels.

Figure \ref{fig:berry_cases}(d) shows the Berry curvature $\Omega_z(k)$ when the antiferromagnetic $\lambda_{AF}=1.95$~meV is applied and the electric-field term is zero $E_z\ell=0$ in this case. It shows that in the $K'$ valley, $|\Delta|$ is small ($\approx 1.95$~meV), so the Berry curvature becomes
large near $k=0$. That is why the red solid curve ($K'\uparrow$) shows a strong positive peak and the red dashed curve ($K'\downarrow$)
shows a strong negative dip. In the $K$ valley, $|\Delta|$ is larger ($\approx 5.85$~meV), so the Berry curvature is much weaker.
That is why the blue solid curve ($K\uparrow$) and the blue dashed curve ($K\downarrow$) stay close to zero compared to the red curves \cite{RevModPhys.82.1959,xiao2007valley,luo2018antiferromagnetic,li2024perfect}. 

Fig. \ref{fig:berry_cases}(e) shows the Berry curvature $\Omega_z(k)$ when $\lambda_{SO}=\lambda_{AF}=3.9$~meV. At the $K'$ valley, the SOC term and the AF term cancel each other, so the gap
becomes zero. Because the gap is closed, the $K'$ valley contribution to the Berry curvature becomes very small, and the red curves stay
almost on the zero line. Meanwhile, the $K$ valley has a huge gap $|\Delta|=7.8$~meV, so its Berry curvature is also
small. That is why the blue curves show only a small peak/dip near $k=0$ and remain close to zero compared to the stronger cases. From the topological point of view, this case represents a phase-boundary situation. 

Figure \ref{fig:berry_cases}(f) shows the Berry curvature $\Omega_z(k)$ when both external fields, the antiferromagnetic exchange is $\lambda_{AF}=3.9$~meV and the electric-field term is $E_z\ell=3.9$~meV are applied. In this regime, the external fields act against the intrinsic spin–orbit coupling, so the Berry-curvature pattern becomes strongly spin and valley dependent. At the $K$ valley, the spin-up branch shows a positive Berry curvature, while the spin-down branch flips sign and becomes negative. By comparison, near the $K$ valley, the Berry curvature remains positive for both spin channels. By comparison, near the $K' $ the Berry curvature remains positive for both spin channels \cite{xu2020topology}.

Figure \ref{fig:magnetic} shows the momentum dependence of the orbital magnetic moment $m_z(k)$ near the $K$ and $K'$ valleys for different values of the perpendicular electric field $E_z$ and antiferromagnetic exchange interaction $\lambda_{AF}$. In general, the orbital magnetic moment exhibits pronounced peaks near $k=0$ corresponding to the vicinity of the Dirac points. This behavior reflects the strong Berry curvature localization around the gapped Dirac points, consistent with previous studies of orbital magnetism in topological materials \cite{xiao2006berry}. In \ref{fig:magnetic}(a), where external stimuli ($E_z=0$ and $\lambda_{AF}=0$) are absent, the system preserves both time-reversal and inversion symmetries except for the intrinsic spin–orbit coupling. The orbital magnetic moments for opposite spin channels have equal magnitude but opposite sign, while the two valleys remain degenerate. This symmetric distribution is a characteristic feature of the quantum spin Hall (QSH) phase, where the system hosts spin-polarized counter-propagating edge states protected by time-reversal symmetry \cite{kane2005quantum,bernevig2006quantum}.

When a finite perpendicular electric field is introduced, as shown in Figure \ref{fig:magnetic}(b) for $E_z=0.5\lambda_{SO}$, inversion symmetry is broken. Consequently, the orbital magnetic moments become valley dependent, and the magnitudes of the peaks at the $K$ and $K'$ points differ for each spin channel. Despite this valley asymmetry, the system remains in the QSH phase because the band gap is still open. At the critical value $E_z=\lambda_{SO}$, shown in Figure \ref{fig:magnetic} (c), the band gap closes for one of the spin–valley sectors, leading to the disappearance of the corresponding orbital magnetic moment. This gap closing signals a topological phase transition from the QSH phase to the quantum valley Hall (QVH) phase.

In Figure \ref{fig:magnetic}(d), with $E_z=0$ and $\lambda_{AF}=0.5\lambda_{SO}$, the exchange field lifts the spin degeneracy and significantly modifies the orbital magnetic moment distribution. Here, the band gap remains finite, and the system retains a QSH-like topological character. When the exchange interaction reaches the critical value $\lambda_{AF}=\lambda_{SO}$, as shown in Figure \ref{fig:magnetic}(e), the Dirac mass vanishes for one spin sector, resulting in another gap-closing point. Similar to the electric-field-driven transition, the orbital magnetic moment associated with the critical band disappears, signaling a topological phase transition.

Figure \ref{fig:magnetic}(f) in this regime, where $E_z=\lambda_{AF}=\lambda_{SO}$, the orbital magnetic moment exhibits strong peaks with the same sign in multiple channels, indicating a net transverse response. This behavior is consistent with the emergence of the quantum anomalous Hall (QAH) phase, characterized by a finite Chern number.

\newpage

\renewcommand{\bibname}{References}

\section*{References}

\bibliography{bib_ref}
\end{document}